# Beyond Log-Supermodularity: Lower Bounds and the Bethe Partition Function

Nicholas Ruozzi
Communication Theory Laboratory
École Polytechnique Fédérale de Lausanne
Lausanne, Switzerland
nicholas.ruozzi@epfl.ch

## Abstract

A recent result has demonstrated that the Bethe partition function always lower bounds the true partition function of binary, log-supermodular graphical models. We demonstrate that these results can be extended to other interesting classes of graphical models that are not necessarily binary or log-supermodular: the ferromagnetic Potts model with a uniform external field and its generalizations and special classes of weighted graph homomorphism problems.

## 1 Introduction

A standard inference problem is to compute the partition function, or normalizing constant, of a given graphical model. As computing the partition function of an arbitrary graphical model is NP-hard, the partition function is often replaced by a more tractable approximation. A popular approximation to the partition function, due to its relationship to the belief propagation algorithm and its practical performance, is given by the Bethe partition function from statistical physics. However, the relationship between the Bethe partition function and the true partition function is difficult to characterize for an arbitrary graphical model.

Using a combinatorial characterization of the Bethe partition function from (Vontobel, 2013), we recently demonstrated that there exist nice families of graphical models for which the Bethe partition function provably lower bounds the true partition function (Ruozzi, 2012). Specifically, for binary graphical models, whenever the potential functions of the graphical model are all log-supermodular, the Bethe partition function always lower bounds the true partition function. As an example, the partition function of the ferromagnetic Ising model with an arbitrary external field can be expressed as the partition function of a log-supermodular function. In a very technical sense, these results can be extended beyond binary graphical models to other models over finite distributive lattices (e.g., any finite totally ordered set) as every finite distributive lattice is isomorphic to a sublattice of the Boolean lattice over $\{0,1\}^n$ for some $n$ (Alon and Spencer, 2000). However, natural candidates for non-binary graphical models, such as the ferromagnetic Potts model, for which one might suspect that the Bethe partition function again provides a lower bound are not log-supermodular in this sense.

In this work, we show that the results of Ruozzi (2012) can be extended to provide bounds on the Bethe partition function of other, not necessarily binary or log-supermodular, graphical models. Specifically, we show that the partition functions of certain graphical models can be equivalently expressed as the partition functions of log-supermodular graphical models over possibly different factor graphs and state spaces. Under certain conditions, this log-supermodular transformation can then be exploited to prove that, again, the Bethe partition function always provides a lower bound on the true partition function.

The models considered in this work include the ferromagnetic Potts model (with some restrictions on the choice of external field), generalizations of the ferromagnetic Potts model to matroids, certain weight enumerators of linear codes, and a subset of graphical models for the weighted graph homomorphism problem. For these models, we demonstrate that the Bethe partition function always provides a lower bound on the true partition function that is provably tighter than the lower bound corresponding to the naïve mean-field partition function.

This paper is organized as follows. In Section 2, we review the relevant background material: graphical models, graph covers, and approximate partition functions. In Section 4, we motivate the results in this work by looking at the simple case of pairwise binary graphical models. In Section 5, we show that the Bethe free energy provides a lower bound on the partition function of the ferromagnetic Potts model with a uniform external field, demonstrating by counter example that the results do not hold for arbitrary external fields. In addition, we show that similar results are true for several common generalizations of the ferromagnetic Potts model that include certain weight enumerators of linear codes. In Section 6, we consider the problem of counting weighted graph homomorphisms and demonstrate that the Bethe partition function provides a lower bound under certain restrictions on the target graph. Finally, we conclude with a short discussion in Section 7.

## 2 Graphical Models

Let $f : \mathcal{X}^n \to \mathbb{R}_{\geq 0}$ be a non-negative function where $\mathcal{X}$ is a finite set. A function $f$ factors with respect to a hypergraph $G = (V, \mathcal{A})$, if there exist potential functions $\phi_i : \mathcal{X} \to \mathbb{R}_{\geq 0}$ for each $i \in V$ and $\psi_\alpha : \mathcal{X}^{|\alpha|} \to \mathbb{R}_{\geq 0}$ for each $\alpha \in \mathcal{A}$ such that

$$f(x_1, \ldots, x_n) = \prod_{i \in V} \phi_i(x_i) \prod_{\alpha \in \mathcal{A}} \psi_\alpha(x_\alpha).$$

The graph $G$ together with the collection of potential functions $\phi$ and $\psi$ define a graphical model that we will denote as $(G; \phi, \psi)$. For a given graphical model $(G; \phi, \psi)$, we are interested in computing the partition function

$$Z(G; \phi, \psi) = \sum_{x \in \mathcal{X}^{|V|}} \Big[ \prod_{i \in V} \phi_i(x_i) \prod_{\alpha \in \mathcal{A}} \psi_\alpha(x_\alpha) \Big].$$

In general, computing the partition function is an NP-hard problem, but in practice, local message-passing algorithms based on approximations from statistical physics, such as loopy belief propagation, produce reasonable estimates in many settings.

### 2.1 Graph Covers

Graph covers have played an important role in our understanding of inference in graphical models (Vontobel, 2013; Vontobel and Koetter, 2005), and in particular, they are intimately related to the approximations of the partition function that we will consider in this work. Roughly speaking, if a graph $H$ covers a graph $G$, then $H$ looks locally the same as $G$.

**Definition 2.1.** *A graph $H$ **covers** a graph $G = (V, E)$ if there exists a graph homomorphism $h : H \to G$ such that for all vertices $i \in G$ and all $j \in h^{-1}(i)$, $h$ maps the neighborhood $\partial j$ of $j$ in $H$ bijectively to the neighborhood $\partial i$ of $i$ in $G$.*

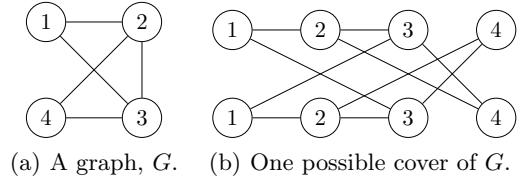

(a) A graph, $G$. (b) One possible cover of $G$.

Figure 1: An example of a graph cover. The nodes in the cover are labeled for the node that they copy in the base graph.

If $h(j) = i$, then we say that $j \in H$ is a copy of $i \in G$. Further, $H$ is said to be an $M$-cover of $G$ if every vertex of $G$ has exactly $M$ copies in $H$. For an example of a graph cover, see Figure 1.

We will typically represent the hypergraph $G = (V, \mathcal{A})$ as a factor graph: a graph with a variable node for each vertex $i \in V$, a factor node for each $\alpha \in \mathcal{A}$, and an edge from $i \in V$ to $\alpha \in \mathcal{A}$ if and only if $i \in \alpha$. In this way, the above definition of graph covers can easily be extended to hypergraphs.

For a connected hypergraph $G = (V, \mathcal{A})$, each $M$-cover consists of a variable node for each of the $M|V|$ variables, a factor node for each of the $M|\mathcal{A}|$ factors, and an edge joining each distinct copy of $i \in V$ to a distinct copy of $\alpha \in \mathcal{A}$ whenever $i \in \alpha$.

To any $M$-cover $H = (V^H, \mathcal{A}^H)$ of $G$ given by the homomorphism $h$, we can associate a collection of potentials: the potential at node $i \in V^H$ is equal to $\phi_{h(i)}$, the potential at node $h(i) \in G$, and for each $\alpha \in \mathcal{A}^H$, we associate the potential $\psi_{h(\alpha)}$. In this way, we can construct a function $f^H : \mathcal{X}^{M|V|} \to \mathbb{R}_{\geq 0}$ such that $f^H$ factorizes over $H$. We will say that the graphical model $(H; \phi^H, \psi^H)$ is an $M$-cover of the graphical model $(G; \phi, \psi)$ whenever $H$ is an $M$-cover of $G$ and $\phi^H$ and $\psi^H$ are derived from $\phi$ and $\psi$ as above.

Finally, it will be convenient to express $f^H$ as a function over $M$ vectors in the set $\mathcal{X}^{|V|}$. We can partition the vertex set $V^H$ into $M$ disjoint sets $V_1, \ldots, V_M$ such that each set contains exactly one copy of each vertex in the graph $G$. Then, without loss of generality, any $x \in \mathcal{X}^{M|V|}$ can be expressed as $x^1, \ldots, x^m \in \mathcal{X}^{|V|}$ where $x^m$ is an assignment to the variables in $V_m$ for all $m \in \{1, \ldots, M\}$. In this case, we will write $f^H(x) = f^H(x^1, \ldots, x^M)$.

### 2.2 The Bethe Approximation

The Bethe free energy is a standard approximation to the so-called Gibbs free energy that is motivated by ideas from statistical physics. At temperature $T = 1$, the Bethe approximation, is defined as follows.

$$\log Z_{\mathrm{B}}(G, \tau; \phi, \psi) = \sum_{i \in V} \sum_{x_i} \tau_i(x_i) \log \phi_i(x_i)$$
$$+ \sum_{\alpha \in \mathcal{A}} \sum_{x_\alpha} \tau_\alpha(x_\alpha) \log \psi_\alpha(x_\alpha)$$
$$- \sum_{i \in V} \sum_{x_i} \tau_i(x_i) \log \tau_i(x_i)$$
$$- \sum_{\alpha \in \mathcal{A}} \sum_{x_\alpha} \tau_\alpha(x_\alpha) \log \frac{\tau_\alpha(x_\alpha)}{\prod_{i \in \alpha} \tau_i(x_i)}$$

for $\tau$ in the local marginal polytope,

$$\mathcal{T} \triangleq \{\tau \geq 0 \mid \forall \alpha \in \mathcal{A}, i \in \alpha, \sum_{x_{\alpha \setminus i}} \tau_\alpha(x_\alpha) = \tau_i(x_i)$$
$$\text{and } \forall i \in V, \sum_{x_i} \tau_i(x_i) = 1\}.$$

The Bethe partition function is defined to be the maximum value achieved by this approximation over $\mathcal{T}$.

$$Z_{\mathrm{B}}(G; \phi, \psi) = \max_{\tau \in \mathcal{T}} Z_{\mathrm{B}}(G, \tau; \phi, \psi)$$

The primary reason for the popularity of this approximation is that fixed points of the belief propagation algorithm correspond to stationary points of $\log Z_{\mathrm{B}}(G, \tau; \phi, \psi)$ over $\mathcal{T}$ (Yedidia et al., 2005). As such, all fixed points of the belief propagation algorithm provide a lower bound on the Bethe partition function. Our goal is to better understand the relationship between the Bethe partition function and the true partition function for different graphical models.

A recent theorem of Vontobel (2013) provides a combinatorial characterization of the Bethe partition function in terms of graph covers.

**Theorem 2.2.**

$$Z_{\mathrm{B}}(G; \phi, \psi) = \limsup_{M \to \infty} \sqrt[M]{\sum_{H \in \mathcal{C}^M(G)} \frac{Z(H; \phi^H, \psi^H)}{|\mathcal{C}^M(G)|}}$$

where $\mathcal{C}^M(G)$ is the set of all $M$-covers of $G$.

*Proof.* See Theorem 33 of (Vontobel, 2013). □

This characterization suggests that bounds on the partition functions of individual graph covers can be used to bound the Bethe partition function. This was the approach taken in (Ruozzi, 2012) and the approach that we will take in this work.

### 2.3 The Naïve Mean-Field Approximation

Another typical approximation, sometimes referred to as the naïve mean-field approximation, is to further restrict the local marginal polytope, the set $\mathcal{T}$ in the definition of the Bethe approximation, so that $\tau_\alpha(x_\alpha) = \prod_{i \in \alpha} \tau_i(x_i)$ for all $\alpha \in \mathcal{A}$.

As it is simply a specialization of $Z_{\mathrm{B}}$, we must have that $Z_{\mathrm{MF}}(G; \phi, \psi) \leq Z_{\mathrm{B}}(G; \phi, \psi)$. However, this does not mean that the mean-field partition function is necessarily a worse approximation to the true partition function $Z(G; \phi, \psi)$, and unlike the Bethe partition function, the mean-field partition function always provides a lower bound on the true partition function (Jordan et al., 1999). More details about the mean-field approximation, its relationship to the Bethe approximation, and some experimental comparisons can be found in (Weiss, 2001).

## 3 Log-supermodularity and Lower Bounds

For a given graphical model $(G; \phi, \psi)$, we are interested in the relationship between the true partition function, $Z(G; \phi, \psi)$, and the Bethe approximation $Z_{\mathrm{B}}(G; \phi, \psi)$. In general, $Z_{\mathrm{B}}(G; \phi, \psi)$ can be either an upper or a lower bound on the true partition function, but for special families of graphical models, $Z_{\mathrm{B}}(G; \phi, \psi)$ is always a lower bound on $Z(G; \phi, \psi)$. In particular, this is true whenever the graphical model consists of only log-supermodular potentials.

**Definition 3.1.** *A function $f : \{0,1\}^n \to \mathbb{R}_{\geq 0}$ is called **supermodular** if for all $x, y \in \{0,1\}^n$*

$$f(x) + f(y) \leq f(x \wedge y) + f(x \vee y) \quad (1)$$

*where $(x \wedge y)_i = \min\{x_i, y_i\}$ and $(x \vee y)_i = \max\{x_i, y_i\}$.*

**Definition 3.2.** *A function $f : \{0,1\}^n \to \mathbb{R}_{\geq 0}$ is called **log-supermodular** if for all $x, y \in \{0,1\}^n$*

$$f(x)f(y) \leq f(x \wedge y)f(x \vee y). \quad (2)$$

**Definition 3.3.** *A graphical model $(G; \phi, \psi)$ is log-supermodular if for all $\alpha \in \mathcal{A}$, $\psi_\alpha(x_\alpha)$ is log-supermodular.*

The definition of submodular and log-submodular functions are obtained by reversing the inequality in (1) and (2) respectively. Log-supermodular functions have a number of special properties: the family is closed under multiplication and marginalization. In addition, they can be maximized in polynomial time. However, in general, computing (or even approximating) the partition function of a log-supermodular graphical model is computationally intractable (Goldberg and Jerrum, 2010).

For any collection of vectors $x^1, \ldots, x^M \in \{0,1\}^n$ let $x^{[1]}, \ldots, x^{[M]}$ denote the collection of vectors such that for all $i \in \{1, \ldots, n\}$ and all $m \in \{1, \ldots, M\}$, $x_i^{[m]}$ is the $m^{th}$ largest element among $x_i^1, \ldots, x_i^M$. Equivalently, if $x^1, \ldots, x^M$ are the columns of some matrix $B$, then $x^{[1]}, \ldots, x^{[M]}$ are the columns of the matrix obtained from $B$ by sorting the elements in each row from greatest to least.

In Ruozzi (2012), the following correlation inequality was proven for log-supermodular functions.

**Theorem 3.4.** *Let $f_1, \ldots, f_M : \{0,1\}^n \to \mathbb{R}_{\geq 0}$ and $g : \{0,1\}^{Mn} \to \mathbb{R}_{\geq 0}$ be nonnegative real-valued functions such that $g$ is log-supermodular. If for all $x^1, \ldots, x^M \in \{0,1\}^n$,*

$$g(x^1, \ldots, x^M) \leq \prod_{m=1}^{M} f_m(x^{[m]}), \qquad (3)$$

*then*

$$\sum_{x^1, \ldots, x^M \in \{0,1\}^n} g(x^1, \ldots, x^M) \leq \prod_{m=1}^{M} \Big[\sum_{x \in \{0,1\}^n} f_m(x)\Big].$$

Applying this theorem to log-supermodular factorizations, with a bit of rewriting, yields the following theorem.

**Theorem 3.5.** *If $(G; \phi, \psi)$ is a log-supermodular graphical model, then for any $M$-cover, $(H; \phi^H, \psi^H)$, of $(G; \phi, \psi)$, $Z(H; \phi^H, \psi^H) \leq Z(G; \phi, \psi)^M$.*

**Corollary 3.6.** *If $(G; \phi, \psi)$ is a log-supermodular graphical model, then*

$$Z_{\mathrm{MF}}(G; \phi, \psi) \leq Z_{\mathrm{B}}(G; \phi, \psi) \leq Z(G; \phi, \psi).$$

As the value of the Bethe approximation at any of the fixed points of BP is always a lower bound on $Z_{\mathrm{B}}(G; \phi, \psi)$, $Z_{\mathrm{B}}(G, \tau; \phi, \psi) \leq Z(G; \phi, \psi)$ for any fixed point of the BP algorithm, with corresponding beliefs $\tau$, as well. Note however, that the inequality between $Z_{\mathrm{B}}(G, \tau; \phi, \psi)$ and the mean-field approximation is only guaranteed to hold for the optimal choice of $\tau$ in the local marginal polytope.

## 4 Pairwise Binary Models

In practice, many graphical models are not naturally formulated as log-supermodular models, but Theorem 3.5 is a statement only about log-supermodular functions. We want to understand when we can use Theorem 3.5 to obtain bounds on $Z_{\mathrm{B}}$ even when the graphical model is not log-supermodular. In this section, we motivate the results in this work by first examining this question for the special case of pairwise binary graphical models (i.e., $\mathcal{X} = \{0,1\}$ and $|\alpha| = 2$ for all $\alpha \in \mathcal{A}$). As each factor of a pairwise model depends exactly two variables, the hypergraph $G = (V, \mathcal{A})$ can be represented as a standard graph, and we will abuse notation and write $G = (V, E)$ in this case.

We can sometimes convert a graphical model that is not log-supermodular into a log-supermodular one by a change of variables (e.g., for a fixed $I \subseteq V$, a change of variables that sends $x_i \mapsto 1 - x_i$ for each $i \in I$ and $x_i \mapsto x_i$ for each $i \in V \setminus I$). These functions are the log-supermodular analog of the "switching supermodular" and "permuted submodular" functions considered in (Crama and Hammer, 2011) and (Schlesinger, 2007) respectively.

To illustrate this idea, consider the special case of log-submodular, bipartite graphical models. A graph $G = (V, E)$ is *bipartite* if the vertex set can be partitioned into two sets $A \subseteq V$ and $B = V \setminus A$ such that $A$ and $B$ are independent sets (i.e., there are no edges between any two vertices $v_1, v_2 \in A$ and similarly for $B$). We will denote bipartite graphs as $G = (A, B, E)$.

Let $G = (A, B, E)$ be a bipartite graph[1] and suppose that $(G; \phi, \psi)$ is a pairwise binary, log-submodular graphical model. For each edge $(a,b) \in E$ with $a \in A$ and $b \in B$ and all $x_a, x_b$, define $\psi'_{ab}(x_a, x_b) \triangleq \psi_{ab}(x_a, 1-x_b)$. Similarly, for each $b \in B$ and all $x_b$, define $\phi'(x_b) \triangleq \phi(1-x_b)$. Finally, for each $a \in A$ and all $x_a$, define $\phi'(x_a) \triangleq \phi(x_a)$. Observe that $\psi'_{ab}$ is a log-supermodular function whenever $\psi_{ab}$ is a log-submodular function. The new graphical model, $(G; \phi', \psi')$ is a pairwise binary, log-supermodular graphical model that is obtained from the original model via a change of variables. Consequently, both graphical models have the same partition functions and the same Bethe partition functions, and we can apply Theorem 3.5 to conclude that $Z(G; \phi, \psi) = Z(G; \phi', \psi') \geq Z_{\mathrm{B}}(G; \phi', \psi') = Z_{\mathrm{B}}(G; \phi, \psi)$.

## 5 The Potts and Random Cluster Models

As a first example of a family of graphical models with non-binary state spaces for which the Bethe partition function provides a lower bound on the true partition function, we consider the Potts model (with no external field) from statistical physics. For a fixed graph $G = (V, E)$, a positive integer $q$, and a vector of spins

---

[1] Factor graphs are always bipartite, but here we are requiring that the graph $G$ is bipartite when represented as a standard graph.

$\sigma \in \{1, \ldots, q\}^{|V|}$,

$$f_{\text{Potts}}^G(\sigma; q, J) = \prod_{(i,j) \in E} e^{J_{ij} \delta(\sigma_i, \sigma_j)}$$

where each edge $(i,j) \in E$ is weighted by $J_{ij} \in \mathbb{R}$ and the notation emphasizes the dependence on the model parameters $J$ and $q$. When $J_{ij} > 0$ for all $(i,j) \in E$, the model is said to be ferromagnetic (i.e., neighboring spins prefer to align), and the model is said to be antiferromagnetic if $J_{ij} < 0$ for all $(i,j) \in E$.

The case $q = 2$ corresponds to the Ising model. Each pairwise potential in the ferromagnetic Ising model is log-supermodular, so we can immediately apply Theorem 3.5 in order to show that $Z_B$ gives a lower bound on the partition function. However, when $q > 2$ the pairwise potential functions need not be log-supermodular, even in the ferromagnetic case.

Our goal in this section is to show that, like switching log-supermodular functions, the ferromagnetic Potts model is also a log-supermodular function in disguise. Unlike the pairwise binary case, we will need more than variable switching in order to produce a log-supermodular function. First, we observe that the partition function of the ferromagnetic Potts model can be equivalently formulated as the partition function of a closely related model in statistical physics, the random cluster model.

The random cluster model is a measure over subsets of edges of the graph $G = (V, E)$ and is related to measures that arise in the study of percolation problems. For any subset $A \subseteq E$ and nonnegative weights $p_{ij}$,

$$f_{\text{rc}}^G(A; q, p) = q^{k_G(A)} \prod_{(i,j) \in A} p_{ij}$$

where $q$ is as above and $k_G(A)$ is the number of connected components of the graph $G = (V, A)$. If $q$ is a positive integer, $J_{ij} \geq 0$ for all $(i,j) \in E$, and $p_{ij} = e^{J_{ij}} - 1$ for all $(i,j) \in E$, then $Z_{\text{rc}}(G; q, p) = Z_{\text{Potts}}(G; q, J)$. A short proof of this standard result from statistical physics can be found in Appendix A, and more details about the Potts model and its relationship to the random cluster model can be found in (Sokal, 2005) and (Grimmett, 2006).

### 5.1 Lower Bounds

As $k_G(A)$ is a supermodular function (here, think of $A$ as being represented by its 0-1 indicator vector), $q^{k_G(A)}$ is a log-supermodular function. From this, we can conclude that $f_{\text{rc}}^G$ is a log-supermodular function whenever $e^{J_{ij}} - 1 > 0$ for all $(i,j) \in E$.

Notice that $f_{\text{rc}}^G$ does not factor over $G$ as $q^{k_G(A)}$ depends on the entire set $A$. In fact, the factor graph corresponding to $f_{\text{rc}}^G$ is a tree: it has one factor node that is adjacent to all of the variable nodes. Although, the Bethe free energy is exact for this model, computing the partition function remains computationally intractable. Instead, we will exploit the theoretical equivalence of the two partition functions to show that for any cover $(H; q, J^H)$ of the ferromagnetic Potts model $(G; q, J)$, $Z_{\text{Potts}}(H; q, J^H) \leq Z_{\text{Potts}}(G; q, J)^M$. Specifically, we will show how to apply Theorem 3.4 directly to the random cluster partition function corresponding to each cover of $(G; q, J)$ in order to show the desired inequality for the Potts model. See the end of Section 2.1 for a reminder of the notation related to graph covers.

**Lemma 5.1.** Let $(H; q, J^H)$ be an $M$-cover of the Potts model $(G; q, J)$. For any $A = (A^1, \ldots, A^M) \subseteq E^H$,

$$f_{\text{rc}}^H(A^1, \ldots, A^M; q, p^H) \leq \prod_{m=1}^{M} f_{\text{rc}}^G(A^{[m]}; q, p)$$

whenever $q \geq 1$ and $p_{ij} = e^{J_{ij}} - 1 \geq 0$ for all $(i,j) \in E^G$.

*Proof.* We will show that for any $M$-cover, $H$, of $G$,

$$k_H(A^1, \ldots, A^M) \leq \sum_{m=1}^{M} k_G(A^{[m]}).$$

The result will then follow from the definition of $f_{\text{rc}}$ and the definition of $H$.

We prove this by induction on $|A|$, the number of edges in the set $A$. For the base case, $|A| = 0$, $k_H(\emptyset) = \sum_{m=1}^{M} k_G(\emptyset)$. Now, suppose $|A| > 0$. Fix one edge in $A = (A^1, \ldots, A^M)$ and let $B = (B^1, \ldots, B^M)$ be the edges of its corresponding connected component. There are two possibilities. First, if $B = A$, then the result follows from the observation that $\sum_{m=1}^{M} k_G(A^{[m]}) \geq \eta - 1 + k_H(A^1, \ldots, A^M)$ where $\eta$ is the maximum element of the vector given by $\sum_{m=1}^{M} A^m$. Otherwise, if $B \neq A$, let $C = A \setminus B$. With the notation that $C = (C^1, \ldots, C^M)$,

$$\begin{aligned} k_H(A) &= k_H(B) + k_H(C) - |V^H| \\ &\stackrel{(a)}{\leq} \sum_{m=1}^{M} k_G(B^{[m]}) + \sum_{m=1}^{M} k_G(C^{[m]}) - |V^H| \\ &\stackrel{(b)}{\leq} \sum_{m=1}^{M} k_G(A^{[m]}) + \sum_{m=1}^{M} k_G(\emptyset) - |V^H| \\ &= \sum_{m=1}^{M} k_G(A^{[m]}) \end{aligned}$$

where (a) follows from the induction hypothesis and (b) follows from the supermodularity of $k_G$. □

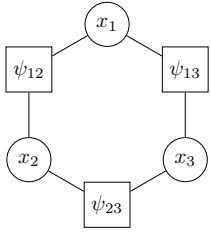

Figure 2: A simple cycle.

**Theorem 5.2.** *For the ferromagnetic Potts model with $q \geq 1$,*

$$Z_{\mathrm{MF}}(G; J, q) \leq Z_{\mathrm{B}}(G; J, q) \leq Z_{\mathrm{Potts}}(G; J, q).$$

*Proof.* By Lemma 5.1 and Theorem 3.4,

$$Z_{\mathrm{rc}}(H; q, p^H) \leq Z_{\mathrm{rc}}(G; q, p)^M$$

for any $M$-cover $(H; q, J^H)$ of $(G; q, J)$ with $p_{ij} = e^{J_{ij}} - 1$ for all $(i,j) \in E^G$ and $p_{ij}^H = e^{J_{ij}^H} - 1$ for all $(i,j) \in E^H$.

The proof then follows from the equivalence between $Z_{\mathrm{Potts}}$ and $Z_{\mathrm{rc}}$ and the characterization of $Z_{\mathrm{B}}$ in Theorem 2.2. $\square$

### 5.2 Potts Models with External Fields

Unlike the results of Ruozzi (2012) for the ferromagnetic Ising model, the results of Section 5.1 do not hold in the case that there is an arbitrary external field (i.e., arbitrary self-potentials). As an example, consider an instance of the ferromagnetic Potts model with an external field that factors over the graph in Figure 2.

$$f(x) = \prod_{k=1}^{3} e^{h_{x_k}^k} \prod_{i \neq j} e^{2\delta(x_i, x_j)}$$

where the vectors $h^1, h^2$, and $h^3$ are defined as follows.

$$h^1 = \begin{bmatrix} e^2 \\ e^{-1} \\ e^{-1} \end{bmatrix} \quad h^2 = \begin{bmatrix} e^{-1} \\ e^2 \\ e^{-1} \end{bmatrix} \quad h^3 = \begin{bmatrix} e^{-1} \\ e^{-1} \\ e^2 \end{bmatrix}$$

For this simple model, we can compute $Z_{\mathrm{B}}$ and $Z$ exactly: $Z_{\mathrm{B}} - Z \approx 973.046$.

In general, the Potts model becomes much more difficult to analyze when there is an external field. However, for uniform external fields, we can extend the results of Section 5.1. Consider the following generalization of the Potts model for a given $h \in \mathbb{R}^q$.

$$f_{\mathrm{Potts}+}^G(\sigma; J, q, h) = \prod_{k \in V} e^{h_{\sigma_k}} \prod_{(i,j) \in E} e^{J_{ij}\delta(\sigma_i, \sigma_j)}$$

The corresponding random cluster representation is given by

$$f_{\mathrm{rc}+}^G(A; q, p, h) = \prod_{C \in \mathrm{comp}(A)} \Big[ \sum_{w=1}^{q} e^{h_w |V(C)|} \Big] \prod_{(i,j) \in A} p_{ij}$$

where $\mathrm{comp}(A)$ is the set of connected components of the graph $(V, A)$ and $V(C)$ is the set of vertices contained in the connected component $C$.

The equivalence between $Z_{\mathrm{Potts}+}$ and $Z_{\mathrm{rc}+}$ for appropriate choices of the parameters can be verified by the same techniques used in Appendix A. As before, $f_{\mathrm{rc}+}^G$ is a log-supermodular function of $A$, though the proof of this statement is non-trivial (see Theorem III.1 of (Biskup et al., 2000)). The analogs of Lemma 5.1 and Theorem 5.2 for this more general model can be proven by using the same arguments as before with minimal modification; we omit the proofs due to space constraints.

**Lemma 5.3.** *Let $(H; q, J^H, h^H)$ be an $M$-cover of the extended Potts model $(G; q, J, h)$. For any $A = (A^1, \ldots, A^M) \subseteq E^H$,*

$$f_{\mathrm{rc}+}^H(A^1, \ldots, A^M; q, p^H, h^H) \leq \prod_{m=1}^{M} f_{\mathrm{rc}+}^G(A^{[m]}; q, p, h)$$

*whenever $q \geq 1$ and $p_{ij} = e^{J_{ij}} - 1 \geq 0$ for all $(i,j) \in E^G$.*

**Theorem 5.4.** *For the extended ferromagnetic Potts model with $q \geq 1$,*

$$Z_{\mathrm{MF}}(G; J, q, h) \leq Z_{\mathrm{B}}(G; J, q, h) \leq Z_{\mathrm{Potts}+}(G; J, q, h)$$

### 5.3 Generalizations to Matroids

Theorem 5.2 and Lemma 5.1 can be extended, with arguments analogous to those in Appendix A, to Potts models defined on hypergraphs as considered in Goldberg and Jerrum (2010). The Potts and random cluster models (like the closely related Tutte polynomial) can also be generalized to matroids Sokal (2005). In this section, we describe the generalization to linear matroids and demonstrate the analog of Lemma 5.1 in this case.

Let $S$ be a matrix over $GF(q)$ for some prime power $q$ whose rows are indexed by the set $V^S$ and whose columns are indexed by the set $\mathcal{A}^S$. The columns of $S$ define a matroid with corresponding rank function $r_S(A)$ which is defined to be the rank, over $GF(q)$, of the submatrix formed by the columns indexed in $A \subseteq \mathcal{A}^S$. For $\sigma \in \{1, \ldots, q\}^{|V^S|}$, the normalized Potts model corresponding to this matroid is

$$f_{\mathrm{Potts}}^S(\sigma; q, J) = \frac{1}{q^{|V^S|}} \prod_{\alpha \in \mathcal{A}^S} \exp \Big[ J_\alpha \delta \big( \sum_{i \in V^S} S_{i,\alpha} \sigma_i, 0 \big) \Big]$$

where $\delta(\sum_i S_{i,\alpha}\sigma_i, 0) = 1$ if $\sum_i S_{i,\alpha}\sigma_i \equiv 0$ over $GF(q)$. Notice that the elements of $S$ form the vertex-edge incidence matrix of a hypergraph $G^S = (V^S, \mathcal{A}^S)$. As a result, we will denote the graphical model for this generalization as $(S; q, J)$.

The partition function of the Potts model for the matroid over $S$ can also be expressed as the partition function of an appropriate generalization of the random cluster model. For each $A \subseteq \mathcal{A}^S$, the normalized random cluster model over $S$ is given by

$$f_{\text{rc}}^S(A; q, p) = q^{-r_S(A)} \prod_{\alpha \in A} p_\alpha.$$

The partition functions of these two models agree whenever $p_\alpha = \exp(J_\alpha) - 1$ for all $\alpha \in \mathcal{A}^S$. When $p_\alpha \geq 0$ for all $\alpha \in \mathcal{A}^S$, $f_{\text{rc}}^S$ is a log-supermodular function. For a proof of the equivalence of $Z_{\text{rc}}(S; q, J)$ and $Z_{\text{Potts}}(S; q, J)$, see Theorem 3.1 of Sokal (2005).

**Lemma 5.5.** *For a prime power $q$ and a matrix $S$ over $GF(q)$, let $(S^H; q, J^H)$ be an $M$-cover of the normalized Potts model $(S; q, J)$.*

*For any $A = (A^1, \ldots, A^M) \subseteq \mathcal{A}^H$,*

$$r_{S^H}(A^1, \ldots, A^m) \geq \sum_{m=1}^M r_S(A^{[m]}).$$

*Proof.* As in the case of Lemma 5.1, the proof follows by induction on $|A| = |A^1| + \cdots + |A^M|$. For $|A| = 0$ or 1, the result follows trivially. Suppose $|A| > 1$. We separate the proof into two cases. In the first case, $r_{S^H}(A) = |A|$. Fix any $a \in A$ and let $B = A \setminus \{a\}$ and $C = \{a\}$. We have

$$r_{S^H}(A^1, \ldots, A^M) = r_{S^H}(B) + r_{S^H}(C)$$
$$\overset{(a)}{\geq} \sum_{m=1}^M r_S(B^{[m]}) + r_S(C)$$
$$\overset{(b)}{\geq} \sum_{m=1}^M r_S(A^{[m]})$$

where (a) follows from the induction hypothesis and (b) follows from the submodularity of $r_S$.

For the second case, $r_{S^H}(A) < |A|$. Choose $B^1 \supseteq \ldots \supseteq B^M$ such that for all $m \in \{1, \ldots, M\}$, $B^m \subseteq A^{[m]}$ and $|B^m| = r_S(A^{[m]})$. This can be accomplished by constructing a basis for the columns of $S$ spanned by $A^{[M]}$ and then extending it to a basis of $A^{[M-1]}$ and so on. Similarly, construct an "unsorted" version of $B^1, \ldots, B^M$ by choosing $C^1, \ldots, C^M$ such that for all $m \in \{1, \ldots, M\}$, $C^m \subseteq A^m$ and $C^{[m]} = B^{[m]}$. With these definitions, we have

$$\sum_{m=1}^M r_S(A^{[m]}) = \sum_{m=1}^M r_S(B^m)$$

$$\overset{(a)}{=} \sum_{m=1}^M r_S(C^m)$$
$$\overset{(b)}{=} r_{S^H}(C^1, \ldots, C^M)$$
$$\overset{(c)}{\leq} r_{S^H}(A^1, \ldots, A^M)$$

where (a) follows from the submodularity of $r_S$, (b) follows from the induction hypothesis, and (c) follows from the monotonicity of $r_{S^H}$. □

With this lemma, we can again use the observation that $Z_{\text{Potts}}$ and $Z_{\text{rc}}$ are equivalent and apply Theorem 3.4 to conclude the following.

**Theorem 5.6.** *For any linear matroid $S$ over $GF(q)$ for some prime power $q$,*

$$Z_{\text{Potts}}(S; J, q) \geq Z_{\text{B}}(S; J, q) \geq Z_{\text{MF}}(S; J, q)$$

*whenever $J_\alpha \geq 0$ for all $\alpha \in \mathcal{A}^S$.*

### 5.4 Weight Enumerators of Linear Codes

As an application of the results of Section 5.3, we demonstrate that the Bethe partition function can be used to compute lower bounds on the weight enumerator of linear codes over $GF(q)$ for some prime power $q$. Similar results have been demonstrated for the problem of counting the number of codewords of a binary cycle code (Vontobel, 2013). Let $S \in GF(q)^{k \times n}$ be the generator matrix of a linear code. A codeword corresponds to any vector in $GF(q)^n$ that is contained in the span of the rows of $S$. Denote the set of valid codewords as $\mathcal{C}$. The weight enumerator of a linear code is defined to be

$$\sum_{c \in \mathcal{C}} \lambda^{w(c)}$$

where $\lambda$ is a positive real and $w(c)$ is the number of nonzero entries in the codeword $c$.

Equivalently, we can formulate this weight enumerator as the partition function of a generalized Potts model.

$$\sum_{c \in \mathcal{C}} \lambda^{w(c)} = \sum_{\sigma \in \{1, \ldots, q\}^k} \lambda^{w(\sum_{i=1}^k \sigma_i S_{i,*})}$$
$$= \sum_{\sigma \in \{1, \ldots, q\}^k} \prod_{\alpha \in \mathcal{A}^S} \lambda^{1-\delta(\sum_{i=1}^k \sigma_i S_{i,\alpha}, 0)}$$
$$= \lambda^{|\mathcal{A}^S|} \sum_{\sigma \in \{1, \ldots, q\}^{|V^S|}} \prod_{\alpha \in \mathcal{A}^S} \lambda^{-\delta(\sum_{i \in V^S} \sigma_i S_{i,\alpha}, 0)}$$
$$= q^k \lambda^n Z_{\text{Potts}}(S; q, \log\frac{1}{\lambda})$$

where $S_{i,*}$ is the $i^{th}$ row of $S$.

Using the previous results for the generalized ferromagnetic Potts model, we have that, for $\lambda \in (0, 1]$,

$$\sum_{c \in \mathcal{C}} \lambda^{w(c)} \geq q^k \lambda^n Z_B(S; q, \log \frac{1}{\lambda})$$

$$\geq q^k \lambda^n Z_{MF}(S; q, \log \frac{1}{\lambda}).$$

As a consequence, belief propagation can be used to compute lower bounds on the weight enumerator in this regime.

## 6 Weighted Graph Homomorphisms

In this section, we consider the problem of counting weighted graph homomorphisms. For a fixed graph $G = (V, E)$, a nonegative matrix $\Gamma \in \mathbb{R}^{n \times n}$, and a vector of nonnegative weights $w \in \mathbb{R}^n$, consider the following function for each $\sigma \in \{1, \ldots, n\}^{|V|}$.

$$f_{\text{hom}}^G(\sigma; w, \Gamma) = \prod_{i \in V} w_{\sigma_i} \prod_{(i,j) \in E} \Gamma_{\sigma_i, \sigma_j}$$

If $\Gamma$ is the adjacency matrix of a graph $\overline{G}$ and $w$ is the all ones vector, then $Z_{\text{hom}}(G; w, \Gamma)$ is equal to the number of graph homomorphisms from $G$ to $\overline{G}$.

We will show that the Bethe partition function provides a lower bound on the true partition function, $Z_{\text{hom}}(G; w, \Gamma)$, whenever

$$\Gamma = aa' + bb', \qquad (4)$$

where $a$ and $b$ are two non-negative column vectors in $\mathbb{R}^n$ and $v'$ denotes the transpose of the vector $v \in \mathbb{R}^n$.

For $\Gamma$ as in (4), $Z_{\text{hom}}(G; w, \Gamma)$ can be reformulated as the partition function of an edge coloring model. For each $A \subseteq E$, define

$$f_{\text{edge}}^G(A; w, a, b) = \prod_{i \in V} \Big[ \sum_{\sigma_i = 1}^n w_{\sigma_i} a_{\sigma_i}^{s_i(A)} b_{\sigma_i}^{|\partial i| - s_i(A)} \Big]$$

where $s_i(A) = |\{j \in \partial i : (i, j) \in A\}|$. By convention, we take $0^0 = 1$.

The proof of equivalence between $Z_{\text{hom}}(G; w, aa' + bb')$ and $Z_{\text{edge}}(G; w, a, b)$ can be found in Appendix B. This proof is a special case of a more general relationship between the partition function of specific vertex coloring functions (our weighted homomorphism function) and the partition function of a more general class of edge coloring functions (Szegedy, 2007).

The edge coloring function $f_{\text{edge}}^G$ is a log-supermodular function. However, unlike the ferromagnetic Potts model and its generalizations, the proof of this observation for the edge coloring model requires some work.

**Lemma 6.1.** *If $a$, $b$, and $w$ are nonnegative vectors in $\mathbb{R}^n$, then $f_{\text{edge}}^G(A; w, a, b)$ is a log-supermodular function of $A \subseteq E$.*

*Proof.* As a product of log-supermodualr functions is log-supermodular, it suffices to show that

$$\sum_{\sigma=1}^n w_\sigma a_\sigma^{s_i(A)} b_\sigma^{|\partial i| - s_i(A)}$$

is log-supermodular for each choice of $i \in V$ or equivalently that

$$\sum_{\sigma=1}^n x_\sigma c_\sigma^{s_i(A)}$$

is a log-supermodular function of $A$ for any nonnegative vectors $x$ and $c$.

Now, observe that, for all $A^1, A^2 \subseteq E$,

$$\Big[\sum_{\sigma=1}^n x_\sigma c_\sigma^{s_i(A^1)}\Big]\Big[\sum_{\sigma=1}^n x_\sigma c_\sigma^{s_i(A^2)}\Big] = \sum_{\sigma, \gamma} x_\sigma x_\gamma c_\sigma^{s_i(A^1)} c_\gamma^{s_i(A^2)}.$$

The desired inequality will follow from the observation that

$$c_\sigma^{s_i(A^1)} c_\gamma^{s_i(A^2)} + c_\sigma^{s_i(A^2)} c_\gamma^{s_i(A^1)} \leq$$
$$c_\sigma^{s_i(A^1 \vee A^2)} c_\gamma^{s_i(A^1 \wedge A^2)} + c_\sigma^{s_i(A^1 \wedge A^2)} c_\gamma^{s_i(A^1 \vee A^2)} \qquad (5)$$

for all $\sigma, \gamma \in [1, \ldots, n]$.

When $\sigma = \gamma$ in (5), the inequality is tight as $s_i(A^1 \wedge A^2) + s_i(A^1 \vee A^2) = s_i(A^1) + s_i(A^2)$. To show the inequality (5) when $\sigma \neq \gamma$, let $s = s_i(A^1 \wedge A^2) + s_i(A^1 \vee A^2)$, and observe that, as a function of $t \in \mathbb{R}$,

$$c_\sigma^t c_\gamma^{s-t} + c_\sigma^{s-t} c_\gamma^t$$

is symmetric about its minimum at $t = s/2$ and monotonic increasing for $t \geq s/2$. The inequality (5) follows from choosing $t = s_i(A^1 \vee A^2) \geq \max\{s_i(A^1), s_i(A^2)\} \geq s/2$.

□

**Theorem 6.2.** *For any graph $G = (V, E)$, any nonnegative $w \in \mathbb{R}^n_{\geq 0}$, and any matrix $\Gamma = aa' + bb'$ for some nonnegative vectors $a, b \in \mathbb{R}^n_{\geq 0}$,*

$$Z_{MF}(G; w, \Gamma) \leq Z_B(G; w, \Gamma) \leq Z_{homs}(G; w, \Gamma).$$

*Proof.* The factor graphs corresponding to the weighted homomorphism model and the edge coloring model are isomorphic (they simply exchange the role of the factor nodes and the variable nodes). As a result, every factor graph corresponding to an $M$-cover of a weighted homomorphism model is isomorphic to a factor graph of some $M$-cover of the edge coloring model. Because, $f_{\text{edge}}^G$ is log-supermodular, an application of Theorem 3.5 gives that $Z_{\text{edge}}(G; w, a, b)^M \geq Z_{\text{edge}}(H; w^H, a^H, b^H)$ for any $M$-cover $(H; w^H, a^H, b^H)$ of $(G; w, a, b)$. The equivalence of the partition functions of the two models combined with the combinatorial characterization of the Bethe partition function in Theorem 2.2 then yields the desired result. □

## 7 Discussion

In a survey of submodular functions, Lovász (1983) noted that "submodularity is not a deep property, but it is often difficult to recognize the circumstances under which it occurs." The above examples serve to illustrate that the same is true of log-supermodularity.

For each model considered in this work, we transformed the problem of computing the partition function of a graphical model $(G; \phi, \psi)$ that is not log-supermodular into the computation of the partition function of a log-supermodular graphical model. This transformation was then applied to graph covers of $(G; \phi, \psi)$ in order to demonstrate that (3) holds for the transformed version of each cover. By the equality of the partition functions, this holds for the partition function of each graph cover of $(G; \phi, \psi)$ as well. Finally, we applied Theorem 3.4 to demonstrate that $Z_B(G; \phi, \psi) \leq Z(G; \phi, \psi)$. This implies that any fixed point of the belief propagation algorithm over the model $(G; \phi, \psi)$ must yield a lower bound on the true partition function.

### Acknowledgments

This work was supported by EC grant FP7-265496, "STAMINA".

## A  The Potts and Random Cluster Models

In this appendix, we present a short algebraic proof of the equivalence between the partition function of the Potts model and the random cluster model under an appropriate choice of parameters. For any graph $G = (V, E)$ and any positive integer $q$, this equivalence follows primarily from the observation that $q^{k_G(A)}$ counts the number of colorings of the vertices of the graph $G$ with $q$ different colors such that any two vertices in the same connected component of the subgraph $(V, A)$ are assigned the same color. This results in the following expression.

$$q^{k_G(A)} = \sum_{\sigma \in \{1,\ldots,q\}^{|V|}} \prod_{(i,j) \in A} \delta(\sigma_i, \sigma_j) \qquad (6)$$

for all $A \subseteq E$.

Given (6), the proof of equivalence follows from simple algebraic manipulations.

$$\sum_{A \subseteq E} f_{\text{rc}}^G(A; q, p) = \sum_{A \subseteq E} \left[ q^{k_G(A)} \prod_{(i,j) \in A} p_{ij} \right]$$
$$= \sum_{A \subseteq E} \left[ \sum_{\sigma \in \{1,\ldots,q\}^{|V|}} \prod_{(i,j) \in A} p_{ij} \delta(\sigma_i, \sigma_j) \right]$$
$$= \sum_{\sigma \in \{1,\ldots,q\}^{|V|}} \left[ \sum_{A \subseteq E} \prod_{(i,j) \in A} p_{ij} \delta(\sigma_i, \sigma_j) \right]$$
$$= \sum_{\sigma \in \{1,\ldots,q\}^{|V|}} \prod_{(i,j) \in E} \left[ 1 + p_{ij} \delta(\sigma_i, \sigma_j) \right]$$

Substituting $p_{ij} = e^{J_{ij}} - 1$ for all $(i,j) \in E$ yields

$$\sum_{A \subseteq E} f_{\text{rc}}^G(A; q, p) = \sum_{\sigma \in \{1,\ldots,q\}^{|V|}} \prod_{(i,j) \in E} e^{J_{ij} \delta(\sigma_i, \sigma_j)}$$
$$= \sum_{\sigma \in \{1,\ldots,q\}^{|V|}} f_{\text{Potts}}^G(\sigma; q, J).$$

## B  The Weighted Homomorphism and Edge Coloring Models

In this appendix, we present a short algebraic proof of the equivalence between the partition function of the weighted homomorphism graphical model and the partition function of the edge coloring model whenever $\Gamma$ is a rank two matrix of the form $aa' + bb'$ for two vectors $a$ and $b$.

$$f_{\text{edge}}^G(A; a, b, w) = \prod_{i \in V} \left[ \sum_{\sigma_i=1}^n w_{\sigma_i} a_{\sigma_i}^{s_i(A)} b_{\sigma_i}^{|\partial i| - s_i(A)} \right]$$
$$= \sum_{\sigma \in \{1,\ldots,n\}^{|V|}} \prod_{i \in V} \left[ w_{\sigma_i} a_{\sigma_i}^{s_i(A)} b_{\sigma_i}^{|\partial i| - s_i(A)} \right]$$
$$= \sum_{\sigma \in \{1,\ldots,n\}^{|V|}} \prod_{i \in V} w_{\sigma_i} \prod_{(i,j) \in E} c_{ij}(A)$$

where

$$c_{ij}(A) = a_{\sigma_i}^{A_{(i,j)}} a_{\sigma_j}^{A_{(i,j)}} b_{\sigma_i}^{1 - A_{(i,j)}} b_{\sigma_j}^{1 - A_{(i,j)}}$$

and $A_{(i,j)} = 1$ if $(i,j) \in A$ and zero otherwise. Notice that $\sum_{A \subseteq E} \prod_{(i,j) \in E} c_{ij}(A) = \prod_{(i,j) \in E} \Gamma_{\sigma_i, \sigma_j}$ and that $c_{ij}(A)$ only depends on whether or not the edge $(i,j) \in A$. Consequently,

$$Z_{\text{edge}}(G; a, b, w) = \sum_{A \subseteq E} f_{\text{edge}}^G(A; a, b, w)$$
$$= \sum_{\sigma \in \{1,\ldots,n\}^{|V|}} \prod_{i \in V} w_{\sigma_i} \prod_{(i,j) \in E} \Gamma_{\sigma_i, \sigma_j}$$
$$= Z_{\text{hom}}(G; w, \Gamma).$$